\documentclass[aps,prl,reprint,superscriptaddress,longbibliography,
floatfix]{revtex4-1}
\usepackage{amsmath,amssymb,graphicx,ulem}
\usepackage[dvipsnames]{xcolor}

\begin{document}

\title{Pairing and pair superfluid density in one-dimensional Hubbard
  models}

\author{B. Gr\'emaud}

\affiliation{Aix Marseille Univ, Universit\'e de Toulon, CNRS,
  CPT, Marseille, France}

\affiliation{MajuLab, CNRS-UCA-SU-NUS-NTU International Joint Research
  Unit, 117542 Singapore}

\affiliation{Centre for Quantum Technologies, National University of
  Singapore, 2 Science Drive 3, 117542 Singapore}

\affiliation{Department of Physics, National University of Singapore,
  2 Science Drive 3, 117542 Singapore}

\author{G. G. Batrouni}

\affiliation{Universit\'e C\^ote d'Azur, INPHYNI, CNRS, 0600 Nice,
  France}

\affiliation{MajuLab, CNRS-UCA-SU-NUS-NTU International Joint Research
  Unit, 117542 Singapore}

\affiliation{Centre for Quantum Technologies, National University of
  Singapore, 2 Science Drive 3, 117542 Singapore}

\affiliation{Department of Physics, National University of Singapore,
  2 Science Drive 3, 117542 Singapore}

\affiliation{Beijing Computational Science Research Center, Beijing
  100193, China}

\begin{abstract}
We use unbiased computational methods to elucidate the onset and
properties of pair superfluidity in two-species fermionic and bosonic
systems with onsite interspecies attraction loaded in one-dimensional
optical lattice. We compare results from quantum Monte Carlo (QMC) and
density matrix renormalization group (DMRG), emphasizing the
one-to-one correspondence between the Drude weight tensor, calculated
with DMRG, and the various winding numbers extracted from the QMC. Our
results show that, for any nonvanishing attractive interaction, pairs
form and are the sole contributors to superfluidity, there are no
individual contributions due to the separate species. For weak
attraction, the pair size diverges exponentially,
i.e. Bardeen-Cooper-Schrieffer (BCS) pairing requiring huge systems to
bring out the pair-only nature of the superfluid.  This crucial
property is largely overlooked in many studies, thereby
misinterpreting the origin and nature of the superfluid.  We compare
and contrast this with the repulsive case and show that the behavior
is very different, contradicting previous claims about drag
superfluidity and the symmetry of properties for attractive and
repulsive interactions.  Finally, our results show that the situation
is similar for soft core bosons: superfluidity is due only to pairs,
even for the smallest attractive interaction strength compatible with
the largest system sizes that we could attain.
\end{abstract}

\maketitle

\underbar{\bf Introduction:} Multi-component quantum systems have long
attracted interest be they bosonic, fermionic or mixtures
thereoff. Andreev and Bashkin\cite{andreev75} considered a two
component mixture of $^4$He and superfluid (SF) $^3$He and showed
that, in addition to the superflows of the indiviual components, there
is a ``drag'' superfluid (DSF) density caused by the repulsive
interaction between the $^4$He atoms and the Cooper pairs formed by
the $^3$He. Interest in such systems increased with the experimental
realization of trapped ultra-cold bosonic and fermionic atoms and
mixtures of the two. The system parameters in such experiments, such
as relative densities and interaction strength, are highly tunable and
can be studied in the bulk or loaded in optical lattices. In this
context, DSF was demonstrated with a microscopic model of a weakly
interacting dilute gas in the bulk\cite{fil05}, and on optical
lattices\cite{linder09,hofer12}. In the bulk or at low densities on a
lattice, the DSF flow is along the SF flow of the separate
components. However, when the coupling is strongly repulsive and the
lattice filling is commensurate, supercounterflow can be
observed\cite{kuklov03,kaurov05,kuklov04a,kuklov04b}. The
two-component DSF density was also studied with mean
field\cite{yanay12,sellin18,hartman18} and quantum Monte Carlo (QMC)
as was the three component case\cite{hartman18}. Mean field gives a
DSF density proportional to the square of the interspecies interaction
for the two-component case leading to the conclusion that this effect
is independent of the sign of the coupling. This would mean that for
both, repulsive and attractive, intercomponent interactions there are
three contributions to the superflow: Those due to the two individual
bosonic components and the DSF component. On the other hand, since
bosons can be more easily cooled to very low temperatures, it was
argued that pairing between fermions can be studied on optical
lattices by considering two-component bosonic system with large
intra-atomic repulsion, mimicking hard core bosons, and attractive
inter-component interaction\cite{paredes03}. It is well known that
attracting fermions undergo pairing correlations: BCS-like for weak
and tight binding (molecular) for strong attraction.  Consequently, in
this picture, all transport is expected to be superconducting (SC):
There is no normal metallic transport when fermions are paired. This
should imply the same behavior in the attractive bosonic case too, but
this is not quite the picture emerging in the recent
literature\cite{hu09,hu11,sellin18}. While there is obvious consensus
that when the attraction is strong enough, the bosons are tightly
paired and transport is only via pair superfluidity (PSF), there is no
consensus on the weakly attractive case. Mean field
calculations\cite{yanay12,sellin18,hartman18} show that for weak
inter-species coupling, the DSF in the repulsive case is equivalent to
the PSF in the attractive case and that the corresponding SF densities
are symmetric with respect to the interaction sign. For bosons with
strong, but finite, intra-species repulsion in one dimension,
renormalization group calculations\cite{hu09} argue that, for weak
interaction, one needs a minimum attraction to form PSF and that PSF
may co-exist with charge density wave (CDW).

We address these issues here and present QMC and DMRG results arguing
that the balanced-population two-component boson system exhibits very
different behavior for positive and negative interspecies
interaction. Specifically, we show that while the repulsive case
displays simultaneous DSF and individual component SF as discussed in
the
literature\cite{andreev75,fil05,linder09,sellin18,hartman18,zhan14,ceccarelli15,ceccarelli16},
the attractive case has only pair SF for both hardcore and softcore
boson systems with no single-component superfluidity. Using the
hardcore boson/fermion analogy, these results also confirm the
similarities in superfluid flow between fermion and boson systems. The
systems we study are one-dimensional because (see below) the sizes
needed to show pairing at weak attractive interaction are very large
and cannot be reached in two dimensions with current algorithms and
computers.

\underbar{\bf Model and methods:} We study two-component Hubbard
models on a one dimensional chain governed by the Hamiltonian
$H=H_0+H_{\text{int}}^{\text{F/B}}$ with
\begin{equation}
\begin{aligned}
H_0&=-t\sum_{i,\sigma} \left (c^{\dagger}_{i,\sigma} c_{i+1,\sigma} +
\mathrm{h.c.}\right )\\ H^{\text{F}}_{\text{int}} &= U \sum_{i}
n_{i,\uparrow}n_{i,\downarrow}\\ H^{\text{B}}_{\text{int}} &= U
\sum_{i} n_{i,\uparrow}n_{i,\downarrow}+U_0 \sum_{i,\sigma}
n_{i,\sigma}(n_{i,\sigma}-1),
\label{hubham}
\end{aligned}
\end{equation}
where F (B) refers to fermions (bosons). The creation (destruction)
operator $c^\dagger_{i,\sigma}$ ($c^{\phantom\dagger}_{i,\sigma}$)
creates (destroys) fermions, soft- or hardcore bosons depending on the
case being discussed. The number operator is $n_{i,\sigma}$. The two
components are labeled by $\sigma=\uparrow,\downarrow$. We fix the
energy scale by taking $t=1$. Hardcore bosons and fermions are related
by the Jordan-Wigner (JW) transformation and share many properties
which we will delineate. 

We studied these models, Eq.(\ref{hubham}), using the ALPS
library\cite{alps} DMRG\cite{white92,white93} with open and
periodic\cite{mondaini18} boundary conditions (OBC, PBC) and the
stochastic Green function (SGF) QMC
algorithm\cite{rousseau08,rousseau08b} with PBC. The OBC DMRG
calculations were done on lattices up to $L=420$, whereas the PBC DMRG
and QMC were done on $L$ up to $120$. In all cases, we verified that the number of DMRG states
and sweeps were sufficient for convergence to the ground state. To
characterize the phases we calculate the single particle and pair
Green functions, $G_{\sigma}(r)$ and $G_p(r)$,
\begin{eqnarray}
\label{singleparticle}
G_{\sigma}(r) &=& \langle c^{\dagger}_{i+r,\sigma} c_{i,\sigma}
\rangle,\\
\label{pairgreen}
G_p(r) &=& \langle P^{\dagger}_{i+r}P_{i} \rangle,\\ 
\label{pairoperator}
P_i &\equiv&c_{i,\uparrow}c_{i,\downarrow},
\end{eqnarray}
where $P_j$ is a pair annihilation operator at site $i$.  Pair
formation is signaled by power law decay of $G_p(r)$ concurrent with
exponential decay of the single particle Green
functions\cite{Giamarchibook,Fradkinbook}, $G_{\sigma}(r)\sim {\rm
  exp}(-r/\xi)$. It is important to note that the JW
  string factors cancel out when computing the pair Green function,
  meaning that fermionic and HCB pair correations always exhibit the
  same behavior in any dimension. In addition, we compute the single
particle charge gap\cite{dodaro17,junemann17},
\begin{equation}
\begin{aligned}
 \Delta=&E(N_{\uparrow}+1,N_{\downarrow})+E(N_{\uparrow}-1,N_{\downarrow})
 -2E(N_{\uparrow},N_{\downarrow})\\ = &
 E(N_{\uparrow},N_{\downarrow}+1) + E(N_{\uparrow},N_{\downarrow}-1)
 -2E(N_{\uparrow},N_{\downarrow}),
\end{aligned}
\label{chargegap}
\end{equation}
where $E(N_\uparrow,N_\downarrow)$ is the ground state energy with
$N_\uparrow$ ($N_\downarrow$) up (down) particles.

We probe transport via the $2\times2$ symmetric Drude weight tensor,
$\mathbf{D}$,
\begin{equation}
 \mathbf{D}_{\sigma\sigma'} = \frac{\pi L}{2t} \frac{\partial^2
   E_0(\Phi_{\sigma},\Phi_{\sigma'})} {\partial
   \Phi_{\sigma}\partial\Phi_{\sigma'}}\bigg
 |_{(\Phi_{\sigma},\Phi_{\sigma'})=(0,0)}.
\end{equation}
The phase twists $\Phi_{\sigma}$ are applied via the replacement
$c_{n\sigma}\to {\rm e}^{in\phi_{\sigma}} c_{n\sigma}$, where
$\phi_{\sigma} =\Phi_{\sigma}/L$ is the phase gradient. This endows
the hopping terms with a phase $\exp{(i\phi_{\sigma})}$ (or its
complex conjugate).  The full tensor $\mathbf{D}$ can be reconstructed
by fitting the curvature of the ground state energy as a function of a
phase $\Phi$ in the following four cases. The single particle weights
are given by the diagonal, $D_{\sigma \sigma}$, calculated with
$(\Phi_{\uparrow},\Phi_{\downarrow})=(\Phi,0)$ or
$(\Phi_{\uparrow},\Phi_{\downarrow})=(0,\Phi)$. The correlated weight
corresponds to applying the same phase gradient on both components,
$(\Phi_{\uparrow},\Phi_{\downarrow})=(\Phi,\Phi)$ giving:
$D=D^{\text{C}}=D_{\uparrow\uparrow}+D_{\downarrow\downarrow}+D_{\uparrow\downarrow}+
D_{\downarrow\uparrow}$. The anti-correlated weight is obtained by
applying opposing gradients on the two components,
$(\Phi_{\uparrow},\Phi_{\downarrow})=(\Phi,-\Phi)$, giving:
$D^{\text{A}} = D_{\uparrow\uparrow} + D_{\downarrow\downarrow} -
D_{\uparrow\downarrow}- D_{\downarrow\uparrow}$.

For bosonic systems these quantities probe superfluid transport and
correspond to the variance of the winding numbers\cite{pollock87}
$W_{\sigma}$,
\begin{eqnarray}
  \label{rhossigma}
\rho_{s\sigma}&=&\frac{L\langle W^2_{\sigma}\rangle}{2t\beta}
=\mathbf{D}_{\sigma\sigma}\\
\label{rhoscorr}
\rho_s^{\text{C}} &=& \frac{L\langle
  (W_{\uparrow}+W_{\downarrow})^2\rangle}{2t\beta}=D^{\text{C}}\\
\label{rhosanticorr}
\rho_s^{\text{A}} &=& \frac{L\langle
  (W_{\uparrow}-W_{\downarrow})^2\rangle}{2t\beta}=D^{\text{A}},
\end{eqnarray}
where $\beta$ is the inverse temperature. This yields
$D_{\downarrow\uparrow}=D_{\uparrow\downarrow}=\frac{L\langle
  W_{\uparrow}W_{\downarrow}\rangle}{2t\beta}$: The off-diagonal term
of {\bf D} and the cross-winding can be used to study directly the
drag and pair SF densities. We calculate
$\mathbf{D}_{\sigma\sigma^\prime}$ using DMRG and for the superfluid
densities we use SGF QMC where windings can be measured directly.

\begin{figure}[h]
  \includegraphics[width=8cm]{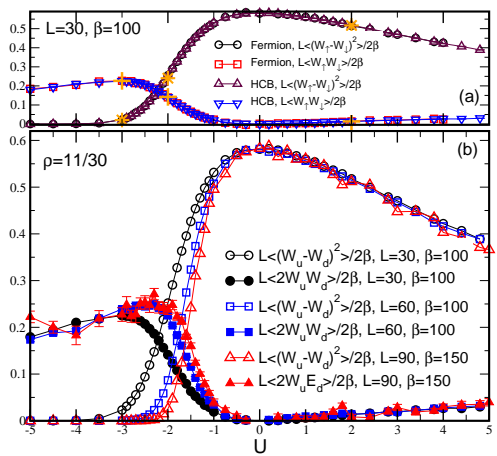}
 \caption{(Color online) (a): Anticorrelated and pair SF density for
   HCB and fermions using SGF QMC show excellent agreement. The plus
   and star (orange) symbols show DMRG results using the Drude weight
   tensor. (b) Finite size dependence of anticorrelated and pair SF
   densities for HCB.}
 \label{hcb-fermions}
\end{figure}

\begin{figure}[h]
 \includegraphics[width=8cm]{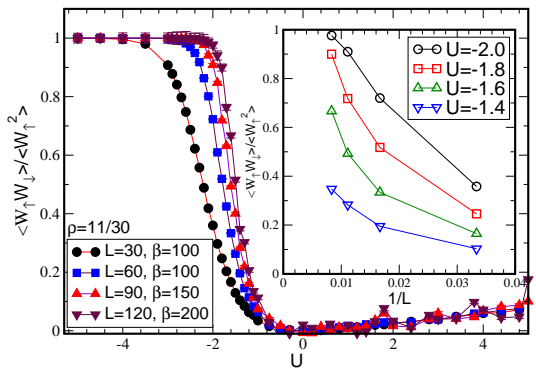}
 \caption{(Color online) The ratio of pair to single component SF
   density rises rapidly to unity for $U<0$. The saturation ratio is
   reached more rapidly as the system size increases. Inset: The same
   ratio as a function of $L^{-1}$ for selected values of $U$. Even
   for the weakest attraction, $U=-1.4$, the ratio increases toward
   saturation as $L$ increases.}
 \label{hcbratios}
\end{figure}

\underbar{\bf Results:} We first verify numerically
Eqs.(\ref{rhossigma},\ref{rhoscorr},\ref{rhosanticorr}) are satisfied
and that results from our QMC and DMRG are in agreement. The top panel
of Fig.\ref{hcb-fermions} shows the anticorrelated and pair SF
densities for fermions and HCB in the range $-5\leq U < 5$. In
addition, we show results for three $U$ values obtained from the Drude
weight tensor using DMRG. We see that agreement is excellent
confirming the coherence of our treatment of fermions and HCB using
QMC and DMRG. The bottom panel of Fig.\ref{hcb-fermions} shows the QMC
evolution of the anticorrelated and pair SF densities with system size
for HCB (and equivalently for fermions) and exhibits some noteworthy
features. When $U>0$, the SF densities suffer very little from finite
size effects.  Also, the DSF ($L\langle W_\uparrow W_\downarrow
\rangle/2\beta$) is small and increases very slowly with $U$. The
situation is different when $U<0$. For large $|U|$, the anticorrelated
SF density vanishes (indicating $\langle W_\uparrow W_\downarrow
\rangle =\langle W_\uparrow^2 \rangle =\langle W_\downarrow^2
\rangle$) and the PSF density suffers very little finite size effects,
indicating that only PSF is present. For small $|U|$, we observe large
finite size effects. The anticorrelated SF density vanishes more
rapidly for larger systems while the PSF density rises more
rapidly. The latter effect suggests that, in the thermodynamic limit,
for any $U <0$, the system is in the PSF phase and all transport is
via pair hopping. This, of course, is expected from the relation
between fermions and HCB via the JW transformation but it emphasizes
the importance of finite size effects for small $|U|$. Another
important feature is the lack of symmetry between $U<0$ and $U>0$, a
feature which persists for soft core bosons contrary to mean field
results\cite{yanay12,sellin18,hartman18}.

\begin{figure}
 \includegraphics[width=9cm]{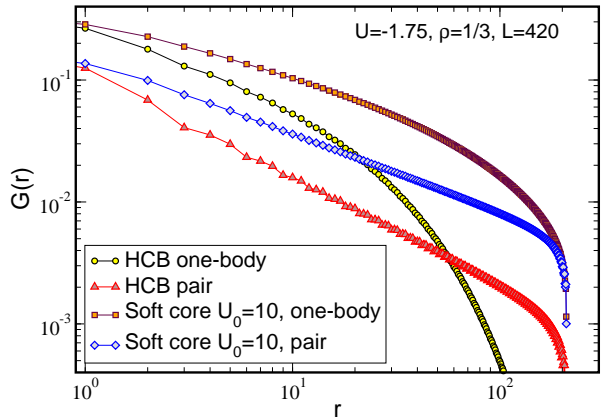}
 \caption{(Color online) Single particle and pair Green functions for
   HCB and soft core bosons ($U_0=10$) at $U=-1.75$. Pair Green
   functions decay as powers whereas the single particle correlations
   decay exponentially for both HCB and soft core bosons. $L=420$ was
   necessary to expose the exponential decay for the soft core case
   due to the much longer correlation length.}
\label{greendecay}
\end{figure}

\begin{figure}[h]
 \includegraphics[width=9cm]{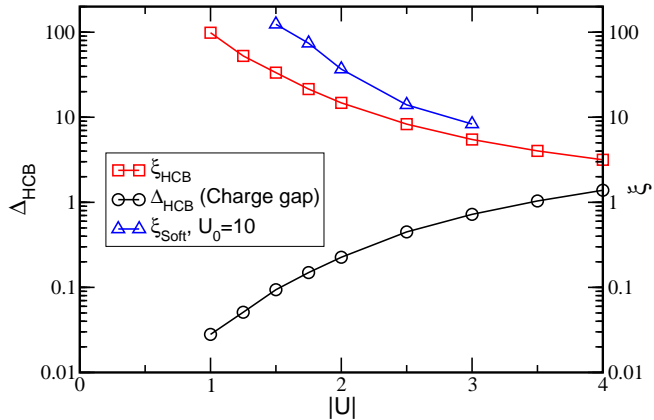}
 \caption{(Color online) Single particle charge gap $\Delta_{HCB}$,
   Eq. \ref{chargegap}, and Green function decay length $\xi_{HCB}$
   for HCB as functions of the $|U|$, at fixed density $n_{\uparrow}=
   n_{\downarrow}= 1/3$. Due to BCS-like behavior at small $|U|$,
   $\xi_{HCB}$ ($\Delta_{HCB}$) diverges (vanishes) exponentially.
   For all $|U|$, the pair gap is always smaller than the numerical
   precision and is consistent with zero.  For soft core bosons at
   $U_0=10$, our results show that $\xi$ exceeds $100$ sites at
   $U=-1.5$, but the accuracy is limited by the
   numerics. Nevertheless, our results cleary emphasize that the
   system is still in a PSF phase for $|U|$ much smaller than
   predicted in ref.\cite{hu09}, see Eq.~\eqref{rgcritu}.}
\label{gapdecay} 
\end{figure}

To elucidate the pair nature of SF when $U<0$, Fig.\ref{hcbratios}
shows, for HCB, the ratio of pair to single species SF densities as a
function of $U$. For $U>0$, this ratio is small (less than $0.1$ for
$U\leq 5$) and insensitive to finite size effects. On the contrary,
when $U<0$, the ratio rises rapidly to unity, and more rapidly the
larger the system. The inset shows this ratio as a function of
$L^{-1}$ for four values of $U$. For $U=-2$, the ratio reaches unity
for $L=120$, but is much smaller for smaller $L$. For $U=-1.8,\,-1.6$,
the ratio does not saturate for the attainable $L$, but it does rise
sharply. For $U=-1.4$ the ratio remains small. However, the mapping to
the fermion system requires this ratio to be unity for any $U<0$. The
reason it is not unity for weak attraction is that this is the regime
of BCS pairing where the correlation length, $\xi$, between the pair
constituents increases exponentially. When $\xi > L$, a false nonzero
single particle SF density will be measured because the members of
such a large pair can still wind around the system independently. DMRG
allows access to larger systems (with OBC) than QMC, and so, we show
in Fig.\ref{greendecay} DMRG results for the single particle and pair
Green functions at $U=-1.75$ and $L=420$. It is clear in this figure
that the pair Green function decays as a power and that the single
particle function decays exponentially. We note, however, that even
though it is unambiguously clear from this figure that, at $U=-1.75$,
only paired HCB participate in superflow, Fig.\ref{hcbratios} shows that the
ratio of pair to single particle SF density is only $0.8$ on a lattice
with $L=120$ sites. The system size needs to be greater than $L=120$
to see the full effect of pairing at this interaction.

In Fig.\ref{gapdecay} we show the growth (decay) of the single
partilce correlation length, $\xi_{HCB}$, (charge gap, $\Delta_{HCB}$)
as $|U|$ gets smaller. We see that for HCB, when $U=-1$, the
correlation length is already $\xi_{HCB} \approx 100$ necessitating
much large system size. We note that the pair gap (energy cost for
adding or removing a pair, not shown in Fig.\ref{gapdecay}) is always
smaller than the numerical precision, as expected for a gapless pair
superfluid.

How these results might extend to soft core bosons (finite $U_0$ in
Eq. (\ref{hubham})) was studied\cite{paredes03} for large $|U_0|$
where they were shown to hold true. Lower values of $|U_0|$ are more
difficult to study because of the exponentially diverging correlation
length. This was addressed\cite{hu09,hu11} numerically and
analytically with the renormalization group. It was shown that, below
full filling, there is critical negative value of the inter-particle
attraction $U$, $U_c$ needed to trigger pair formation,
\begin{equation}
  \frac{U_c}{U_0} = -32 \frac{t^2}{U_0^2}{\rm sin}^2(\pi \rho),
  \label{rgcritu}
\end{equation}
where $\rho$ is the particle density of each species. For $U_c < U
<0$, pairs do not form and the system is in a phase made of an equal
mixture of two independent superfluids. For $U<U_c$, a PSF phase is
established which may even co-exist with CDW. We investigated these
claims using DMRG with $U_0=10$, and $\rho= 1/3$ and system sizes up
to $L=420$. Equation (\ref{rgcritu}) then predicts $U_c = -2.4$. Our
DMRG results for the one-body and pair Green functions, for soft
bosons at $U=-1.75$, are presented in Fig.\ref{greendecay} and show
clearly that pair correlations decay as a power law while one-body
correlations decay faster (exponential). In Fig.\ref{gapdecay} we show
the decay length, $\xi_{Soft}$, down to $|U|=-1.5$ where the
correlation function is still exponential, and $\xi_{Soft}\approx
100$. Even at this very small value, the system is still in the PSF
phase whereas it is predicted to be\cite{hu09} in a mixture of two
independent SF. Note that, even though our results do not definitively
exclude the possibility that the PSF phase does terminate at a much
smaller but finite negative $U$, we see no evidence for this. This
agrees with Ref.~\cite{paredes03}, where the overlap between the exact
soft-core boson and HCB ground states was shown to be close to one and
a smooth function of $U$.

\underbar{\textbf{Conclusion:}} In summary, we have shown that, for
HCB (and fermions) in 1D optical lattice, a BCS-like quantum phase
transition, in the BKT universality class \cite{Giamarchibook}, takes
place as soon as an attractive interaction ($U<0$) exists: the system
is driven into a PSF phase where single particle transport is fully
suppressed and superfluidity is due only to pairs.  As we
demonstrated, for $U<0$ the BCS-like exponential growth of the
one-body Green function decay length, $\xi_{HCB}$, has been widely
overlooked leading to misinterpreted finite-size driven transport
properties. This is in sharp contrast with the repulsive case ($U>0$)
where transport comprises both a single particle and a two body
component (drag superfluid), i.e. where power-law decays are much less
sensitive to finite size. We expect a similar pairing behavior to hold
for higher dimensions, i.e. a BCS-like transition to pair
superfluidity as soon as $U<0$ for HCB, making finite size effects
even more of a limiting issue for numerical simulations.

Finally, for soft bosons, our results show that down to rather weak
attractive interaction, the transport properties are very similar to
those of HCB \textbf{and fermions.} We emphasize that, if
it exists, the phase made of two independent superfluids and no
pairing, would be present only for a much narrower interaction range
than predicted in Ref.~\cite{hu09}. A more thorough study of this
problem is beyond the scope of this paper and would require both a
much more extended computational effort and a revised renormalization
group analysis.

\underbar{\bf Acknowledgments:} The computations were performed with
resources of the National Supercomputing Centre, Singapore
(www.nscc.sg). This research is supported by the National Research
Foundation, Prime Minister's Office and the Ministry of Education
(Singapore) under the Research Centres of Excellence programme.

\acknowledgments

\end{document}